\documentclass[sigconf]{acmart}
\AtBeginDocument{%
  \providecommand\BibTeX{{%
    \normalfont B\kern-0.5em{\scshape i\kern-0.25em b}\kern-0.8em\TeX}}}
\usepackage{xspace}
\usepackage{amsthm}
\usepackage{amsmath}
\usepackage{enumitem}
\usepackage{booktabs}
\usepackage{multirow}
\usepackage{graphicx}
\usepackage[normalem]{ulem}
\useunder{\uline}{\ul}{}
\usepackage{caption}
\usepackage{subcaption}
\usepackage{balance}
\newcommand{\modelname}{\textsf{MSACL}\xspace}

\setlist[itemize]{leftmargin=*}

\setcopyright{acmcopyright}
\copyrightyear{2018}
\acmYear{2018}
\acmDOI{10.1145/1122445.1122456}

\acmConference[Woodstock '18]{Woodstock '18: ACM Symposium on Neural
  Gaze Detection}{June 03--05, 2018}{Woodstock, NY}
\acmBooktitle{Woodstock '18: ACM Symposium on Neural Gaze Detection,
  June 03--05, 2018, Woodstock, NY}
\acmPrice{15.00}
\acmISBN{978-1-4503-XXXX-X/18/06}

\begin{document}

\title{Episodes Discovery Recommendation with Multi-Source Augmentations}

\author{Ziwei Fan}
\authornote{Work is done during the internship in Spotify Research.}
\email{zfan20@uic.edu}
\affiliation{%
  \institution{University of Illinois Chicago}
  \country{USA}
  \postcode{43017-6221}
}

\author{Alice Wang, Zahra Nazari}
\email{{alicew, zahran}@spotify.com}
\affiliation{%
  \institution{Spotify Research}
  \country{USA}}

\renewcommand{\shortauthors}{Trovato and Tobin, et al.}

\begin{abstract}
Recommender systems~(RS) commonly retrieve potential candidate items for users from a massive number of items by modeling users' interests based on historical interactions. However, historical interaction data is highly sparse, and most items are long-tail items, which limits the representation learning for item discovery.
This problem is further augmented by the discovery of novel or cold-start items. For example, after a user displays interest in bitcoin financial investment shows in the podcast space, a recommender system may want to suggest \textit{e.g.,} a newly released blockchain episode from a more technical show. 
Episode correlations help the discovery, especially when interaction data of episodes is limited. Accordingly, we build upon the classical Two-Tower model and introduce the novel \textbf{M}ulti-\textbf{S}ource \textbf{A}ugmentations using a \textbf{C}ontrastive \textbf{L}earning framework~(\modelname) to enhance episodes embedding learning by incorporating positive episodes from numerous correlated semantics. Extensive experiments on a real-world podcast recommendation dataset from a large audio streaming platform demonstrate the effectiveness of the proposed framework for user podcast exploration and cold-start episode recommendation.
\end{abstract}

\maketitle

\section{Introduction}
Recommender Systems~(RS) have been widely applied to numerous web~\cite{lin2020fill, lin2022phish} applications to retrieve relevant information. 
Collaborative filtering~(CF), as the most commonly used RS~\cite{koren2022advances,he2017neural,zheng2019gated,liu2022federated}, assumes that users with similar interests prefer similar items. The users' interests are typically modeled and optimized by historical interactions~\cite{rendle2010factorization,koren2009matrix,10.1145/3459637.3482242,yang2022large}. However, as CF-based RS models interest based on historical interactions, these methods can only capture interests observed in training data and fails to explore topics that users might be interested in but may never know. Therefore, a significant challenge for RS is to facilitate user exploration~\cite{chen2021values}. Exploration is increasingly becoming a critical problem in RS, as existing RS methods can cause \textit{echo chambers} and \textit{filter bubbles} as users increasingly engage with RS~\cite{ge2020understanding}. These two coined terms~\cite{rowland2011filter,allen2017effects} introduce a phenomenon where users' interests become self-reinforced, as only items matching with users' past interests are exposed to users by RS. This phenomenon optimizes short-term user interests and fails to drive long-term user engagement. Moreover, the lack of diversity of recommended items also potentially reduces user satisfaction. We build an RS for podcast recommendation that promotes exploration and podcast discovery here.

In recent years, podcast listening~\cite{nazari2020recommending} has shown a tremendous increase in popularity\footnote{https://www.edisonresearch.com/the-infinite-dial-2021-2/}. This growth in user interest, as well as a relatively low barrier to creation compared to other media such as music or movies, has created an explosion of podcast creation. Therefore, new and diverse podcast content is increasingly and continuously being created.
We argue that the bottleneck of the user podcast exploration problem for RS consists of two challenges, including \textit{feature sparsity} and \textit{interaction sparsity}~\cite{fan2022sequentiala,fan2021modeling}. These two challenges come from the data sparsity problem in RS~\cite{grvcar2005data, 10.1145/3404835.3463036}, where limited interaction data is available for representing users and items. The reason for feature sparsity is that many podcast contents are cold-start items with few user interactions. 
Moreover, in the problem of user exploration, there is a lack of user-item interaction training data. 

To resolve these challenges, we propose a novel framework with interaction-level data augmentations from multi-sources episodes correlation semantics~\cite{liu2020basket} in contrastive learning~(\modelname), building upon the classical backbone of various recommender systems, Two-Tower architecture~\cite{huang2013learning}. Data augmentation~\cite{wei2021contrastive,yu2022graph,xie2022contrastive} enriches data with different views of similar items for learning item embeddings, and contrastive learning~\cite{tian2020makes,khosla2020supervised,zimmermann2021contrastive} acts as a bridge connecting augmented items and positively interacted items.
The proposed \modelname framework proposes the discovery item augmentation and the discovery contrastive regularization to alleviate the two challenges. For the discovery item augmentation, we incorporate similar items from different semantic item relationships~\cite{10.1145/3459637.3482092,fan2022sequential} as positive items to enrich scarce user-item interactions. The discovery contrastive regularization further connects the user-item predictions with the similarity between discovery item and positive item. Our contributions are summarized as follows:
\begin{itemize}
    \item A novel data augmentation framework \modelname is proposed to alleviate data sparsity in user exploration and recommendation for episode discovery from two perspectives: (1). discovery augmentation of different semantic similarities, and (2). discovery contrastive regularization for alleviating user-item explorative interactions sparsity with augmented discovery items.
    \item Extensive experiments on a large, real-world podcast recommendation dataset demonstrate the effectiveness of the proposed framework \modelname for episode explorative recommendation, especially for cold episodes.
\end{itemize}


\section{\modelname}
\subsection{Problem Definition}
In RS, we denote the user set and episode set as $\mathcal{U}$ and $\mathcal{I}$, in which each user and episode are denoted as $u$ and $i$. For an episode $i$, it belongs to one podcast show $p$ but one podcast show has multiple episodes $\{i\in p\}$, where $p\in\mathcal{P}$ and $\mathcal{P}$ denotes the set of podcast shows. The interactions between users and episodes are represented as a set $\{r_{ui}\in\mathcal{R}\}$, where $r_ui$ is binary indicating whether the user $u$ has listened to the episode $i$ with more than 30 seconds. For each user and episode, we also have feature vectors $f_{u}\in\mathbb{R}^{d_{f_{u}}}$ and $f_{i}\in\mathbb{R}^{d_{f_{i}}}$, where $d_{f_{u}}$ and $d_{f_{i}}$ are dimension sizes of the user input feature vector and the episode feature vector. The episode exploration recommendation is predicting the preference score of the user $u$ to an episode $i$ that is out of user $u$'s historical interests, \textit{i.e.,} $i\not\in\{p|r_{uj}=1, j\in p,\text{for all } j\in\mathcal{I}\}$:
\begin{equation}
\label{eq:two_tower_pred}
    \hat{r_{ui}} = F_{\text{u}}(f_u)^\top F_{\text{i}}(f_i),
\end{equation}
where $F_{\text{u}}$ and $F_{\text{i}}$ are neural networks for learning user embeddings and episode embeddings. We generate the episode exploration recommendation list by ranking the scores $r_{ui}$ on all episodes in descending order.

\subsection{Two-Tower Model for Recommendation}
The Eq.~(\ref{eq:two_tower_pred}) describes the standard Two-Tower architecture for recommendation~\cite{huang2013learning}, which comprises of the user tower~($F_{\text{u}}$) and the episode tower~($F_{\text{i}}$). To be specific, the $L$-th layers fully connected neural network output embeddings of user $u$ and episode $i$ are as follows:
\begin{equation}
\label{eq:tower_model}
    \begin{aligned}
        F_\text{u}^L &= \text{ReLU}(F_\text{u}^{L-1}W^L_{\text{1}} + b^L_{\text{1}})\\
        F_\text{\text{i}}^L &= \text{ReLU}(F_\text{i}^{L-1}W^L_{\text{2}} + b^L_{\text{2}}),
    \end{aligned}
\end{equation}
where $\text{ReLU}(\cdot)$ refers to the ReLU activation function, the 0-th layer of each tower is the input feature vector of $f_u$ or $f_i$, respectively, $W^L_{*}\in\mathbb{R}^{d_{L-1}\times d_L}$ are linear transformations and $b^L_{*}\in\mathbb{R}^{d_L}$ are bias. The last layer's output embeddings will be used to make predictions as in Eq.~(\ref{eq:two_tower_pred}). 

\subsection{Discovery Items Augmentation}
The scarcity of user-episode interactions is a significant issue of recommendation for discovery and exploration. The Two-Tower architecture in Eq.~(\ref{eq:two_tower_pred}) demands feature interactions to generalize to unseen user-episode discovery recommendations. However, feature interactions based on only training interaction data are limited due to the data scarcity of user-episode interactions. To alleviate this issue, we extract more positive episodes from additional item similarity semantic relationships, including episodes with similar text content and episodes with similar knowledge information. We assume that a user will be more likely to explore novel episodes with similar content or correlated knowledge to items they have interacted with in the past. Note that discovery items may have similar semantics in knowledge and contents but are not significant in feature interactions. Each item has a large number of features in practical scenarios, and semantic information is diminished. 
Comparing directly using content and knowledge embeddings as features, 
augmenting similar episodes from different semantics provides more information as augmented episodes have other features and also enrich feature interactions in the Two-Tower architecture.

Specifically, for each episode $i$, we have pre-trained content embeddings~\cite{reimers2019sentence} and knowledge embeddings~\cite{yang2014embedding} as side information. The content embeddings are pre-trained with the episode script and title text, and the knowledge embeddings are pre-trained from the episode knowledge graph data. We adopt the Approximate Nearest Neighbors lookup with Annoy to extract the top-K similar episodes from each semantic relationship, which we denote them as $S_{\text{content}}(i)=\{j\in \text{Annoy}_{\text{content}}(K)\}$, and $S_{\text{kg}}(i)=\{j\in \text{Annoy}_{\text{kg}}(K)\}$. We extract the top-K similar episodes by ranking the top-K episodes with smallest L2 distances on content embeddings or knowledge embeddings, respectively. We use top-10 for simplicity. 

\subsection{Discovery Contrastive Regularization}
With the augmented discovery episodes from multiple sources of semantics, we introduce the contrastive learning loss for enriching feature interactions and alleviating the data sparsity in both features and instances.
We can create `positive' episodes $i$ from similar semantic relationships~($\{i_{+}\in S_{\text{content}}(i)\}$ or $\{i_{+}\in S_{\text{kg}}(i)\}$), even the combination of both augmentations. To be specific, given a minibatch of $N$ user-episode exploration interactions $\{(u_k, i_k)\}_{k=1}^N$, we augment one `positive' episode~(augmented episode) $i^{u_k}_{+}$ for the $k$-th interaction. In total, we have $2N$ episodes and we reindex all episodes, we obtain:
\begin{equation}
    \{i^{u_1}_a, i^{u_1}_b, i^{u_2}_a, i^{u_2}_b, \cdots, i^{u_N}_a, i^{u_N}_a,\},
\end{equation}
and the $(i^{u_k}_a,i^{u_k}_b)$ is the positive pair of the user $k$, $a$ and $b$ subscripts denote the positive item $a$ and the semantically similar item $b$ for the user $k$, and other $2N-1$ pairs are considered as negative pairs. For each episode pair $(i^{u_k}_a,i^{u_k}_b)$, their features are $(f_{i^{u_k}_a}, f_{i^{u_k}_b})$, we obtain the learned embeddings $F_{i^{u_k}_a}^L$, $F_{i^{u_k}_b}^L$, and the learned user embedding $F_{u_k}^L$ after $L$-layers of episode tower layer defined in Eq.~(\ref{eq:two_tower_pred}). We adopt the NT-Xent loss~\cite{fang2020cert,gao2021simcse,chen2020simple,he2020momentum} for optimization as follows:
\begin{equation}
\label{eq:cl_loss}
    \mathcal{L}_{CL} = -\log\frac{\text{exp}\left({F_{i^{u_k}_a}^L}^\top F_{i^{u_k}_b}^L\right)}{\sum_{m=1}^{2N-1}\text{exp}\left({F_{i^{u_k}_a}^L}^\top F_{i_m}^L\right)}.
\end{equation}
Note that we maximize dot product between the positive and augmented episodes $(F_{i^{u_k}_a}^L, F_{i^{u_k}_b}^L)$, which matches with the user-episode exploration prediction defined in Eq.~(\ref{eq:two_tower_pred}). With the combination of the dot product between the positive and augmented episodes in Eq.~(\ref{eq:cl_loss}) and the dot product between the user and episode interaction prediction in Eq.~(\ref{eq:two_tower_pred}), we can conclude that it is equivalent to augment user-episode interactions with more `positive' $(u_k, i^{u_k}_b)$ interactions.

With different variants of augmentations, we investigate several models in this framework, including the model with only feature dropout~TT-FD, the content similar augmentation~\modelname(Content), the knowledge similar augmentation~\modelname(KG), and also the combination of feature dropout and knowledge similar augmentation~\modelname(KG-FD). We adopt the sampled softmax cross-entropy loss function with $k$ sampled negative items as the user-episode interactions optimization and also incorporate the contrastive loss as follows:
\begin{equation}
\label{eq:cl_loss}
    \mathcal{L} = -\sum_{(u, i)\in\mathcal{R}}[\log (\sigma(\hat{r_{ui}})) + \sum_{j=1}^k\log(1-\hat{r_{uj}})] + \lambda \mathcal{L}_{CL},
\end{equation}
where $\lambda$ is a hyper-parameter for controlling the contribution from $\mathcal{L}_{CL}$.

Note that we maximize dot product between the positive and augmented episodes $(F_{i^{u_k}_a}^L, F_{i^{u_k}_b}^L)$, which matches with the user-episode exploration prediction defined in Eq.~(\ref{eq:two_tower_pred}). With the combination of the dot product between the positive and augmented episodes in Eq.~(\ref{eq:cl_loss}) and the dot product between the user and episode interaction prediction in Eq.~(\ref{eq:two_tower_pred}), we can conclude that it is equivalent to augment user-episode interactions with more `positive' $(u_k, i^{u_k}_b)$ interactions.

With different variants of augmentations, we investigate several models in this framework, including the model with only feature dropout~(TT-FD), the content similar augmentation~(TT-Corr(Content)), the knowledge similar augmentation~(TT-Corr(KG)), and also the combination of feature dropout and knowledge similar augmentation~(TT-FD-Corr(KG)).

\subsection{Loss and Optimization}
We adopt the sampled softmax cross-entropy loss function as the user-episode exploration interaction optimization and also incorporate the NT-Xent contrastive loss as regularization, which are summarized as follows:
\begin{equation}
    \mathcal{L} = -\sum_{(u, i)\in\mathcal{R}}[\log (\sigma(\hat{r_{ui}})) + \sum_{j=1}^k\log(1-\hat{r_{uj}})] + \lambda \mathcal{L}_{CL},
\end{equation}
where we sample $k$ negative items in interactions optimization. 

\section{Experiments}
In this section, we present the experimental settings and results to demonstrate the effectiveness of \modelname. We answer the following Research Questions~(RQs):
\begin{itemize}
    \item \textbf{RQ1}: Is \modelname effective in episode exploration recommendation?
    \item \textbf{RQ2}: How does each correlation relationship contribute to the performance?
    \item \textbf{RQ3}: Does \modelname achieve better performance in cold items?
\end{itemize}

\subsection{Dataset Statistics and Preprocessing}
\begin{table}[]
\centering
\caption{Dataset Statistics After Preprocessing}
\label{tab:data_stat}
\begin{tabular}{@{}cccc@{}}
\toprule
Data Part & \#users & \#episodes & \#interactions \\ \midrule
Train & 65,756 & 78,103 & 77,003 \\
Validation & 8,230 & 17,381 & 9,642 \\
Test & 8,248 & 17,758 & 9,551 \\ \bottomrule
\end{tabular}
\end{table}
We conduct experiments on data collected from a large audio streaming platform. In this dataset, users are recommended podcast episodes.  Here, our work focuses on exploration.  We define discovery as user-episode interactions from podcast shows the user has never interacted with. Note that each podcast show may have multiple episodes, but we impose a stringent definition of discovery; user interactions with novel episodes from familiar shows are not considered discovery. We define positive interactions to be episode listening of at least 30 seconds. To this end, the dataset contains a sample of 82,234 users on 113,242 episodes with 96,196 user exploration interactions, and the density is 0.001\%. We separate users into training, validation, and test sets via the split ratio of 8:1:1 as users and items have features as inputs. Detailed data statistics are listed in Table~\ref{tab:data_stat}. The features we used for the user tower includes gender, age, country, podcast topics liked in previous 90 days, user language, pre-trained collaborative filtering embedding vector,  the user embedding pre-trained with podcast interactions, and averaged streaming time. For episodes, we use features including topics, country, collaborative filtering pre-trained embeddings, and pre-trained semantic embeddings of the podcast. We use two different pre-trained semantic embeddings, which we call KG (knowledge graph) and Content.  The KG embeddings are learned by applying a common KG embedding method, DistMult~\cite{yang2014embedding}, on a graph that contains available metadata on the podcasts such as episode, topic, licensor, and publisher nodes.  Edges are created between episode-licensor, episode-topic, episode-publisher, and topic-topic.  The Content embeddings are obtained from using pre-trained BERT embeddings on the podcast titles and descriptions ~\cite{reimers2019sentence}.  For discrete features, we encode them as one-hot or multi-hot vectors. For continuous features, such as pre-trained embedding vectors, we directly use them as inputs. After encoding, we concatenate all features of users and items, respectively.

\subsection{Baselines}
As the proposed framework \modelname builds upon a Two-Tower model, we compare with two groups of baselines, standard popularity-based methods and two-tower based methods. We use three popularity-based methods.  One is Pop method, which ranks items based on the number of positive interactions. The other two popularity-based methods are Pop-Country and Pop-Age-Country, which rank items' number of interactions conditioned on the users' country and the combination of age and country, respectively. For Two-Tower-based methods, the first baseline is the standard Two-Tower~(TT) model with the full feature set. The second is the Two-Tower method with feature dropout~(TT-FD) proposed by~\cite{yao2021self}. For parameter setting, we search the number of layers of Two-tower models from $\{1,2,3\}$, the dropout probability from $\{0.1, 0.3, 0.5, 0.7\}$.

\subsection{Evaluation Metrics}
We evaluate the proposed \modelname on standard ranking metrics for top-N recommendation, including Recall@N, NDCG@N, and MRR. Recall@N measures the percentage of positive items being recommended in top-N ranking lists. NDCG@N considers the positions of correctly ranked positive items in top-N recommendation lists. Mean Reciprocal Rank~(MRR) is similar to NDCG@N with also the consideration of ranking positions but in the entire ranking list. We report the averaged evaluation metrics over all test users.

\subsection{Overall Comparison~(RQ1)}
\begin{table}[]
\centering
\caption{Performance Comparison in Recall@10, NDCG@10, Recall@20, NDCG@20, and MRR. The best and second-best results are boldfaced and underlined, respectively.}
\label{tab:overall_perf}
\resizebox{0.48\textwidth}{!}{%
\begin{tabular}{@{}cccccc@{}}
\toprule
Model & Recall@10 & NDCG@10 & Recall@20 & NDCG@20 & MRR \\ \midrule
Pop & 0.02349 & 0.01325 & 0.03618 & 0.01681 & 0.01220 \\
Pop-Country & 0.03115 & 0.01839 & 0.05239 & 0.02430 & 0.01727 \\
Pop-Age-Country & 0.02898 & 0.01701 & 0.04707 & 0.02198 & 0.01560 \\
TT & 0.06068 & 0.04039 & {\ul 0.08594} & 0.04726 & 0.03732 \\
TT-FD & {\ul 0.06153} & {\ul 0.04131} & 0.08548 & {\ul 0.04816} & {\ul 0.03784} \\
\modelname & \textbf{0.06609} & \textbf{0.04336} & \textbf{0.09062} & \textbf{0.05018} & \textbf{0.03966} \\ \midrule
Improv vs TT & +8.92\% & +7.36\% & +5.45\% & +6.18\% & +6.28\% \\
Improv vs 2nd-best & +7.41\% & +4.96\% & +5.45\% & +4.19\% & +4.81\% \\ \bottomrule
\end{tabular}
}
\end{table}
We report the overall results with all baselines in Table~\ref{tab:overall_perf}. We have the following observations:
\begin{itemize}
    \item The proposed method \modelname achieves the best performance over the second best baseline from 4.19\% to 7.41\% on all metrics. Compared with TT, the improvements are from 5.45\% to 8.92\%. These observations demonstrate the superiority of \modelname for episode exploration and recommendation. The reasons for the improvements are twofold: (1). the feature and instance level data augmentation alleviates the features and interactions data sparsity problem for episode exploration and recommendation; (2). the incorporation of additional positive items from multiple semantics of item similarities.
    \item Among TT-based methods, we can see that \modelname performs the best. TT-FD~\cite{yao2021self} achieves the second best performance in most metrics. TT-FD outperforms TT model, therefore demonstrating the utility of contrastive learning in episodes recommendation. We also observe that \modelname achieves better performance than TT-FD. This observation supports the superiority of instance augmentation in contrastive learning.
    \item We can also see that the TT baseline method significantly outperforms the popularity baselines, which demonstrates the effectiveness of incorporating features into the users and items modeling.
    \item Among popularity-based baselines, we can observe that the baseline using users' country information achieves the best performance. The second-best one is using both age and country information. The reason why using age information degrades the performance is that the combination of age and country will separate users' interactions in extremely fine-grain buckets, where most items have no interaction.
\end{itemize}

\subsection{Effects of Different Correlations~(RQ2)}
\begin{table}[]
\centering
\caption{Performance Comparisons for Different Item Similarity Semantics. The best and second-best results are boldfaced and underlined, respectively.}
\label{tab:different_semantics}
\resizebox{0.48\textwidth}{!}{%
\begin{tabular}{@{}cccccc@{}}
\toprule
Model & Recall@10 & NDCG@10 & Recall@20 & NDCG@20 & MRR \\ \midrule
TT-FD & 0.06153 & 0.04131 & 0.08548 & 0.04816 & 0.03784 \\
\modelname(Content) & 0.06088 & 0.04045 & 0.08430 & 0.04721 & 0.03708 \\
\modelname(KG) & {\ul 0.06385} & {\ul 0.04272} & {\ul 0.08792} & {\ul 0.04952} & {\ul 0.03909} \\
\modelname(FD-KG) & \textbf{0.06609} & \textbf{0.04336} & \textbf{0.09062} & \textbf{0.05018} & \textbf{0.03966} \\ \bottomrule
\end{tabular}
}
\end{table}
We incorporate positive episodes from different correlation semantics in \modelname to enhance the instance augmentation. We investigate different types of correlations, including two ways of computing semantic similarity between episodes (one is content-based similarity and another is knowledge-based similarity~\cite{10.1145/3459637.3481934}). The feature dropout also creates a noisy version of positive episodes, providing an alternative view of episodes, which can also be viewed as similar episodes. We show the performance results of different correlations in Table~\ref{tab:different_semantics}. We obtain the following observations:
\begin{itemize}
    \item The combination of feature dropout and knowledge graph-based instance augmentation perform the best among all variants. We can conclude that both feature level and instance level augmentations are necessary for the episodes exploration recommendation.
    \item Comparing \modelname(KG) and \modelname(Content), we can observe that knowledge graph-based similar correlations identify more informative positive episodes for matching users' interests in exploration. This is reasonable that knowledge embeddings encode more heterogeneous episodes relationships while content-similar episodes might bring limited unexpected episodes to users that are also of interest.
\end{itemize}

\subsection{Cold Start Items Performance~(RQ3)}
\begin{figure} 
     \centering
     \begin{subfigure}[b]{0.235\textwidth}
         \centering
        \includegraphics[width=1\textwidth]{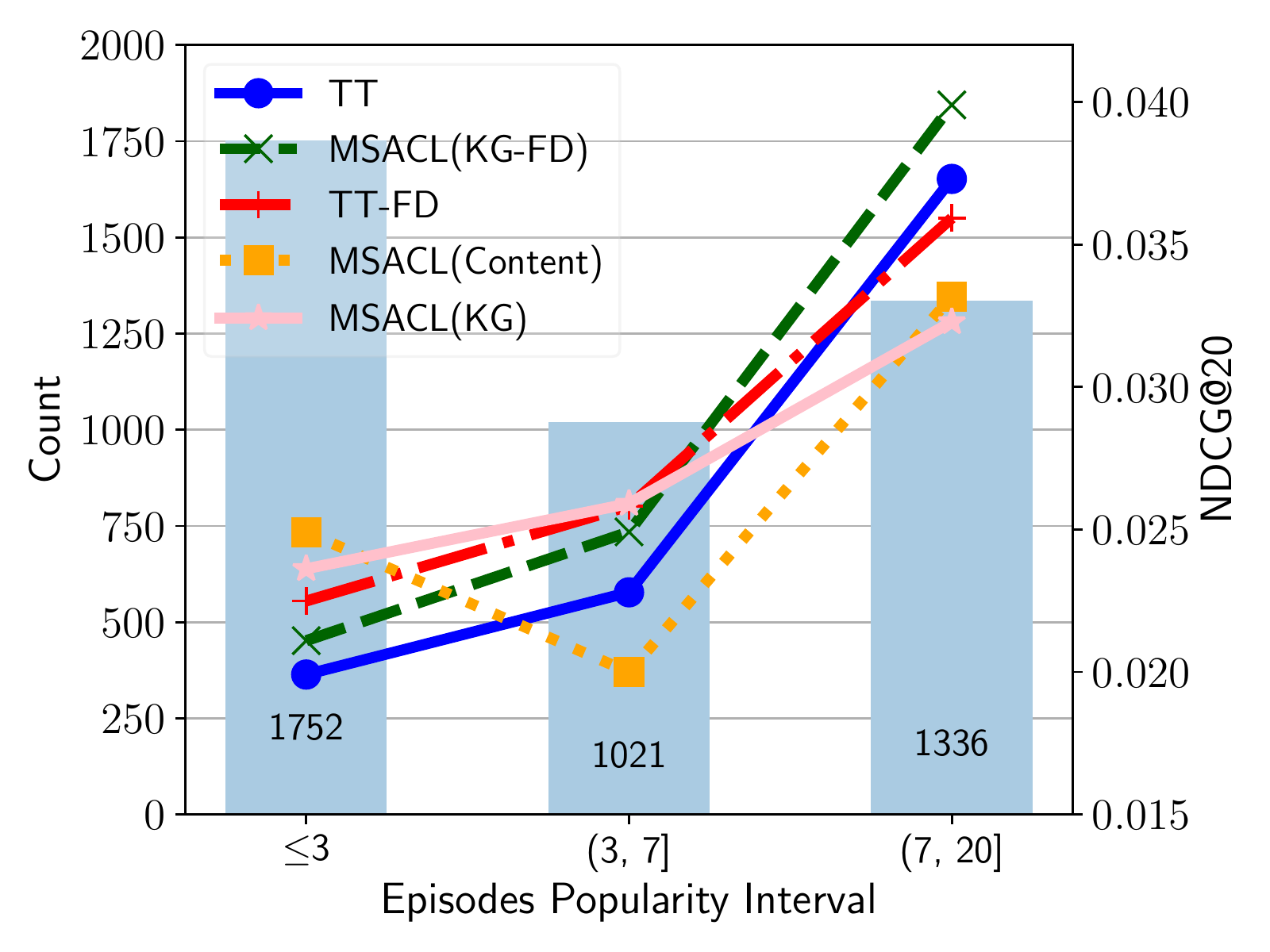}
        \caption{Cold Episodes}
        \label{fig:spotify_unpopular_item}
     \end{subfigure}
     \begin{subfigure}[b]{0.235\textwidth}
         \centering
        \includegraphics[width=1\textwidth]{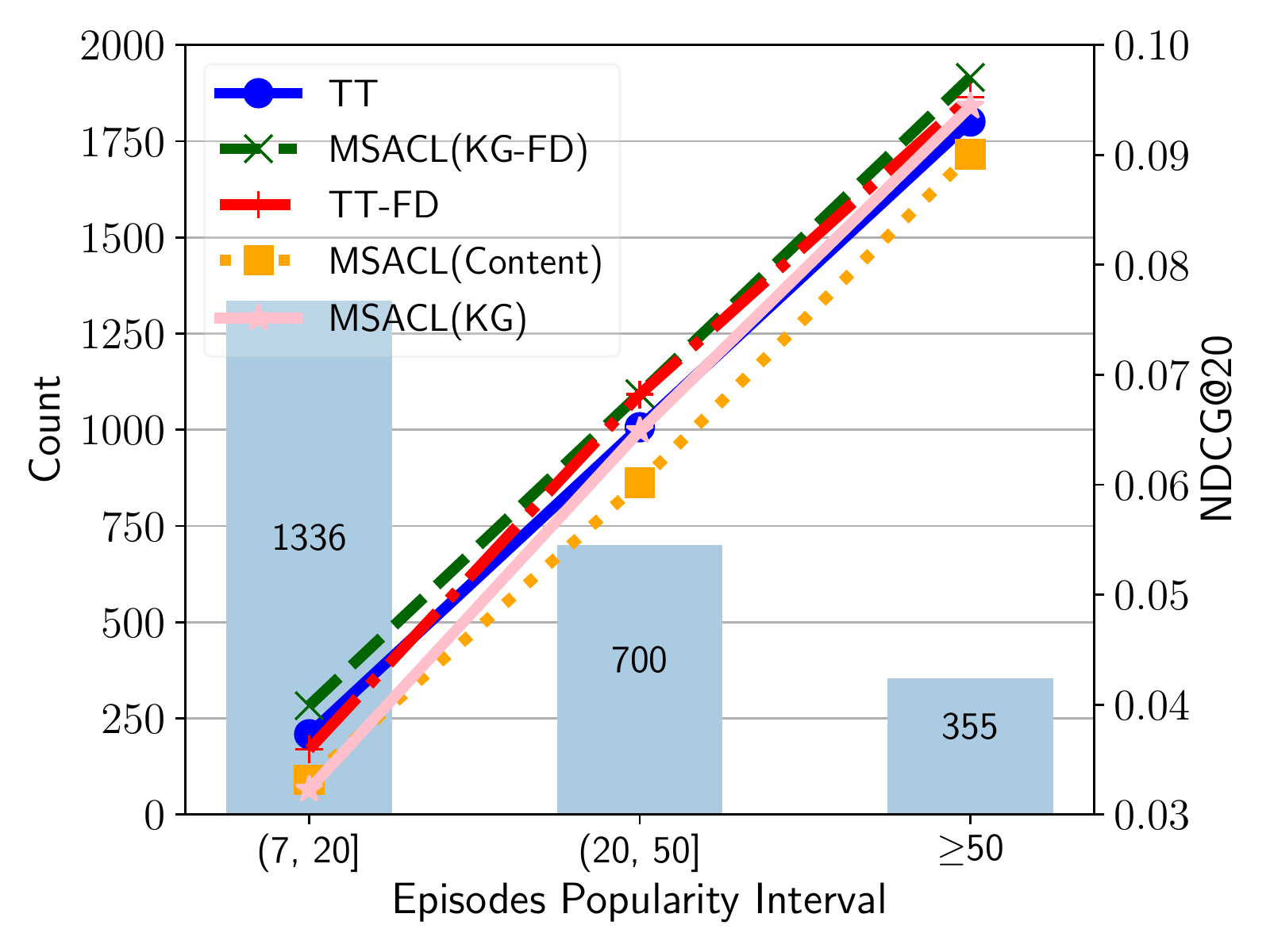}
        \caption{Popular Episodes}
        \label{fig:spotify_popular_item}
     \end{subfigure}
    \vspace{-5mm}
     \caption{The NDCG@20 performance on items with different popularity.}
     \label{fig:perf_wrt_item_pop}
\end{figure}
A major performance bottleneck for the episode exploration and recommendation problem is due to data sparsity, which our framework \modelname can alleviate. We separate episodes into groups based on popularity and visualize the average performance of each group, as shown in Figure~\ref{fig:perf_wrt_item_pop}. We visualize cold items in Figure~\ref{fig:spotify_unpopular_item} and popular items in Figure~\ref{fig:spotify_popular_item}. We can first observe that cold episodes comprise the majority while popular episodes are in the minority. Next, we have additional two observations:
\begin{itemize}
    \item For cold episodes recommendation, all variants of the proposed framework \modelname obtain significant improvements on extremely cold episodes~(with no more than three interactions). It demonstrates the superiority of the proposed model \modelname with contrastive learning and hierarchical augmentations for user exploration recommendation. The \modelname(KG-FD) achieves the most consistent improvements in all cold episodes group, which indicates the necessity of both feature level and instance level augmentations.
    \item Regarding popular episodes, we can see that all methods, including the baseline method, have the same trend, which is that performance increases as popularity increases and that all methods achieve similar performance on the most popular episodes.
\end{itemize}

\section{Conclusion}
In this work, we propose a novel data augmentation framework to solve a podcast exploration recommendation problem. To alleviate challenges regarding feature and instance sparsity, we propose a contrastive learning framework using feature and instance augmentations and achieve improvements over state-of-the-art baselines under the Two-Tower architecture. We also demonstrate the effectiveness of the proposed framework for cold episode recommendation in the increasingly influential domain of podcasts.

\bibliographystyle{ACM-Reference-Format}
\balance
\bibliography{ref}

\end{document}